\documentclass[fleqn,12pt]{wlscirep}
\usepackage[utf8]{inputenc}
\usepackage[T1]{fontenc}
\usepackage{ulem}
\usepackage{graphicx}
\usepackage{amsmath}
\usepackage{amsfonts}
\usepackage{amssymb}
\usepackage{subfigure}
\usepackage{amsthm}
\usepackage{amssymb}
\usepackage{amsxtra}     
\usepackage{epsfig}
\usepackage{verbatim}
\usepackage{graphics}
\usepackage{wrapfig}
\usepackage{subfigure}
\usepackage{float}
\usepackage{morefloats}
\usepackage{xcolor}
\usepackage{optidef}
\usepackage{algorithmicx}
\title{Origami-Inspired Composite Springs with Bi-directional Translational-Rotational Functionalities }

\author[1,*]{Ravindra Masana}
\affil{Research Associate, Laboratory of Applied Nonlinear Dynamics (LAND), New York University Abu Dhabi, UAE.}
\author{ Mohammed F. Daqaq}
\affil[2]{Global Network Professor, Tandon School of Engineering, New York University, Brooklyn, NY.}

\affil[*]{rm4829@nyu.edu}

\begin{abstract}

Many of the patterns seen in Origami are currently being explored as a platform for building functional engineering systems with versatile characteristics that cater to niche applications in various technological fields. One such pattern is the Kresling pattern, which offers unconventional mechanical properties with rich coupled translation and rotational kinematics. In this paper, we design and manufacture a composite spring inspired by the Kresling Origami pattern, which is capable of simultaneously behaving as an axial and torsional restoring element with bi-directional functionalities. We study, numerically and experimentally, the restoring behavior of this spring, its equilibria, and their bifurcations for different combination of the design parameters. We show that the fabricated springs can have fixed, quasi-zero, or variable stiffness, and can be customized to exhibit single- or multi-stable states (symmetric and asymmetric) as needed. The proposed spring demonstrates how combining additive manufacturing with Origami principles can offer a new pathway towards the design of new structural and machine elements with versatile functionalities.
\end{abstract}
\begin{document}

\flushbottom
\maketitle
\section{Introduction}

Origami is the art of folding paper to create aesthetically pleasing three-dimensional designs. Long before its practice as a craft  \cite{MahadevanOrigami2005,Faber1386}, forms of origami existed in nature \cite{ArthurOrigamiReview2014,SchenkReview2014,CAVALLO2015538}, and were recently uncovered by various researchers who turned to the fields of biology and physiology of plants and animals to gain further insights into building multi-functional engineering systems \cite{Janine1997Biomimicry,PereraReviewBiomimicry,DUPLESSIS2019BiomimeticAM,Li_2017}. The appearance of origami patterns in nature inspired such researchers to explore origami as a platform for building functional engineering systems with versatile characteristics that cater to niche applications in various technological fields \cite{SchenkReview2014,TurnerReviewOrigami2016,Sli2019_ArchitectedOrigami,Jiang2014Origami}. This includes the design and construction of structures with auxeticity \cite{schenk2013geometry}, multi-stability \cite{waitukaitis2015origami}, and programmable stiffness \cite{Silverberg2014,fang2018programmable}. Such structures have already found their way into the design of solar arrays \cite{MiuraOrigami1985}, inflatable booms \cite{SchenkReview2014}, vascular stents \cite{KURIBAYASHI2006131}, viral traps \cite{MONFERRER2023101237,Dey2021DNA}, wave guides \cite{Jiang2014Origami,OverveldeOrigamiMetamaterials2016,FilipovMetaStructures2015,BabaeeOrigamiWaveguide2016},  and robotic manipulators \cite{FeltonOrigamiMachine2014,DanielaReviewOrigamiRobot2018}.


Among the many different available origami patterns, some designs have attracted more attention in engineering applications. For example, the \textit{Miura-Ori}, a rigid origami pattern\footnote{In rigid origami structures only  creases exhibit deformation during deployment.}, has been used to construct three-dimensional deployable structures that have been studied and utilized in applications including space exploration~\cite{MiuraOrigami1985,nishiyama2012miura,LangOrigami}, deformable electronics \cite{SongOrigamiBatteries2014,Rui2014}, artificial muscles \cite{OrigamiMuscles2017}, and reprogrammable mechanical metamaterials \cite{Silverberg2014,Sengupta2018,Sli2019_ArchitectedOrigami}. The \textit{Ron Resch}, which is a non-periodic rigid origami with unusually high buckling strength has been used for energy absorption, \cite{Jiang2014Origami,Chen2018EnergyAbsorption}.

The \textit{Yoshimura} pattern \cite{Yoshimura1955,Reid2017Origami} and the \textit{Kresling} pattern \cite{Kresling2008,Butler2016,RmasanaKIOS19,Kidambi2020,kaufmann2020harnessing,Kidambi2020Kresling} are leading examples of non-rigid origami patterns \footnote{Non-rigid origami structures exhibit deformation of the panels between the creases during deployment. This incorporates other degrees of freedom (hidden degrees of freedom) that are free from the kinematic constraints that govern the rigid folding \cite{SilverbergNature2015}} which have been utilized to engineer structures with unique properties.  For instance, the \textit{Kresling} pattern, has inspired the design of flexible tunable antennas \cite{Yao_OrigamiAntenna2014_2}, robot manipulators \cite{Twistedtowerrobot}, wave guides \cite{yasuda2019origami}, selectively-collapsible structures \cite{ZhaiKresling2018}, vibration isolators \cite{ishida2017design},  fluidic muscles \cite{OrigamiMuscles2017}, mechanical bit memory switches \cite{YasudaKresling2017,MasanaKIMS,jules2021delicate}, reconfigurable antenna\cite{ZHANG2022107470} and crawling and peristaltic robots \cite{TawfickKreslingRobot2017,Ruike2022Robotcrawler,BHOVAD2019100552}.

\begin{figure*}[htb!]

                               \centering

                               \includegraphics[width=14cm]{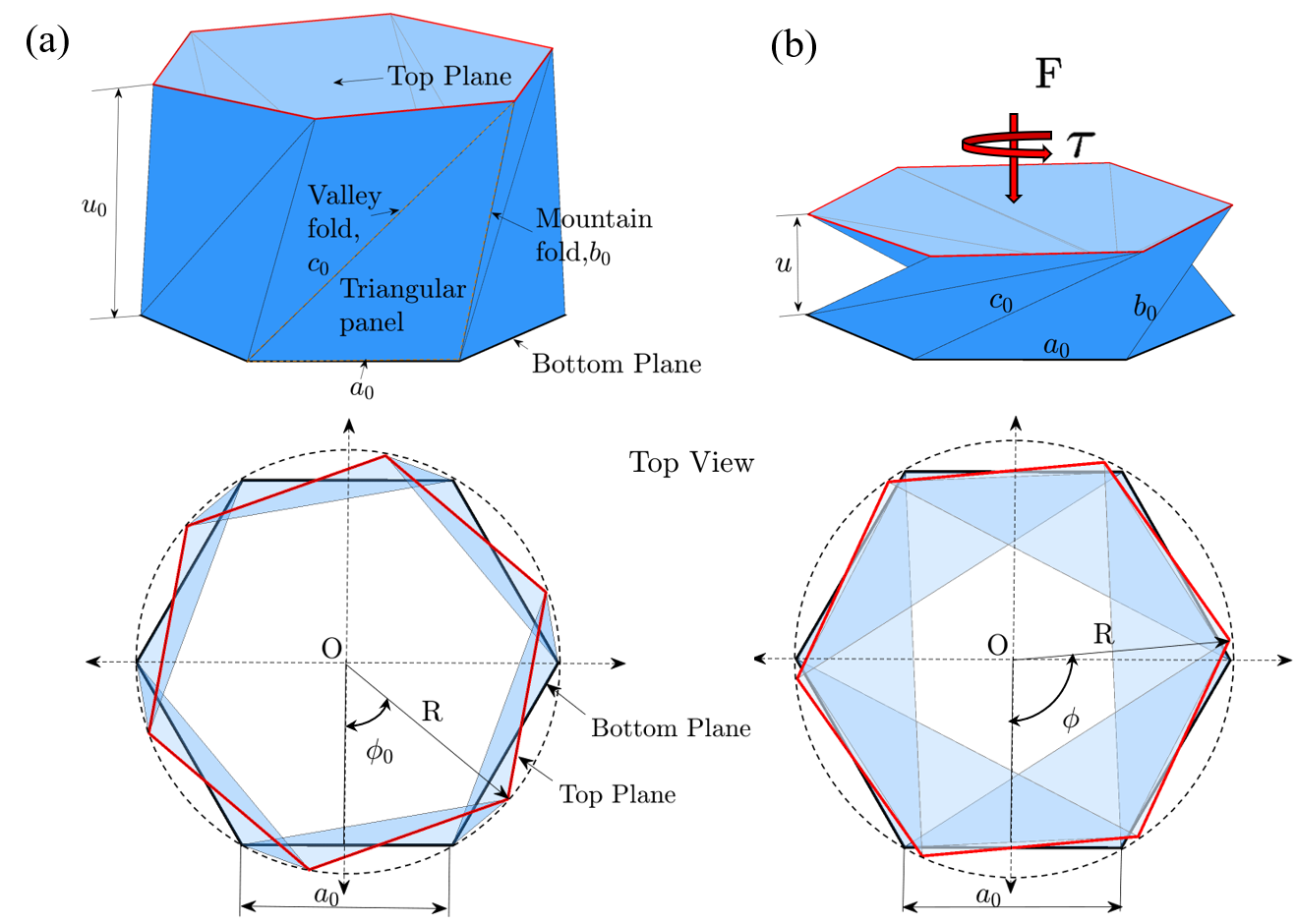}

                               \caption{(a) Schematic representation of the KOS with $n=6$ polygon, and (b) schematic representation of the operation of KOS under applied load.} 

                               \label{Scheme}

               \end{figure*}

The Kresling pattern has also been used to build and construct coupled linear-torsional springs coined as Kresling Origami Springs (KOSs) \cite{KHAZAALEH2022109811}. Such springs which take the shape of a cylindrical bellow-type structure are created by tessellating similar triangles in cyclic symmetry and connecting them as shown in Figure \ref{Scheme}(a). The triangles in the KOS are connected in a circular arrangement, with each triangle connected to two other triangles along two of its edges. One edge forms a mountain fold, $b_0$, and the other a valley fold, $c_0$. The third edges, $a_0$, of the connected triangles form two parallel polygonal end planes (top and bottom planes). The design of the KOS is characterized by geometric parameters, that include the number of sides, $n$, of the parallel polygons, the radius, $R$, of the circle that encloses them, the preloading height, $u_0$, and rotation angle, $\phi_0$ between the end planes. Figure \ref{Scheme}(a) illustrates these parameters.

When a Kresling Origami Spring (KOS) is subjected to an axial load or a torque, it undergoes compression or expansion, depending on the direction of the load. As a result, the two parallel polygon planes, while staying rigid, move and rotate relative to each other along a centroidal axis, as shown in Fig. \ref{Scheme}(b). This motion causes the triangular panels to deform and store energy in the form of strain energy. Upon removal of the external load, the KOS springs back to its initial configuration, releasing the stored energy.

The behavior of this Origami-inspired KOS is that of a unique restoring element with axial and torsional functionalities, which can form the foundation for many exciting engineering structures, especially those in the field of rotating machinery machinery\cite{Lee20221}, haptics\cite{BAO2022104119}, and soft robotics\cite{BHOVAD2019100552,Ruike2022Robotcrawler}. However, because of the nature of its coupled kinematics, a single KOS always results in a coupled translational-rotational motion regardless of the type of load applied to it. In other words, the two coordinates,  $u$, and, $\phi$, are always kinematically constrained resulting in a single degree of freedom. Thus, when one end of the KOS is fixed while the other (free end) is subject to a load, either axial or torsional, the free end undergoes coupled translational-rotational motion. This coupled motion of the free end is not desirable since, in most applications, the free end is usually constrained from rotating when the load is axial, and from translating when the load is torsional.

As such, it is desired that the two motions of the free end be decoupled so that the restoring force is independent for different types of loads. One way to achieve this goal is to join two KOSs end-to-end in series to form a Kresling Origami Spring Pair (KOSP). Unlike a single KOS, where the two coordinates of the free end are always coupled, the KOSP can either have coupled or decoupled motion, depending on the angle of the creases in its constituent KOSs. When the constituent KOSs are joined in a way that their creases have opposite sign slopes with respect to the horizontal connecting surface, the motion at the free end is decoupled. On the other hand, if the constituent KOSs are connected in a way that their creases have similar sign slopes, motion at the free end remains coupled, but with an extended range of operation.   For brevity, we will refer to KOSPs with decoupled motion at the free end as $d$-KOSPs, while those with coupled motion will be denoted as $c$-KOSPs.

\begin{figure*}[htb!]

                               \centering

                               \includegraphics[width=16cm]{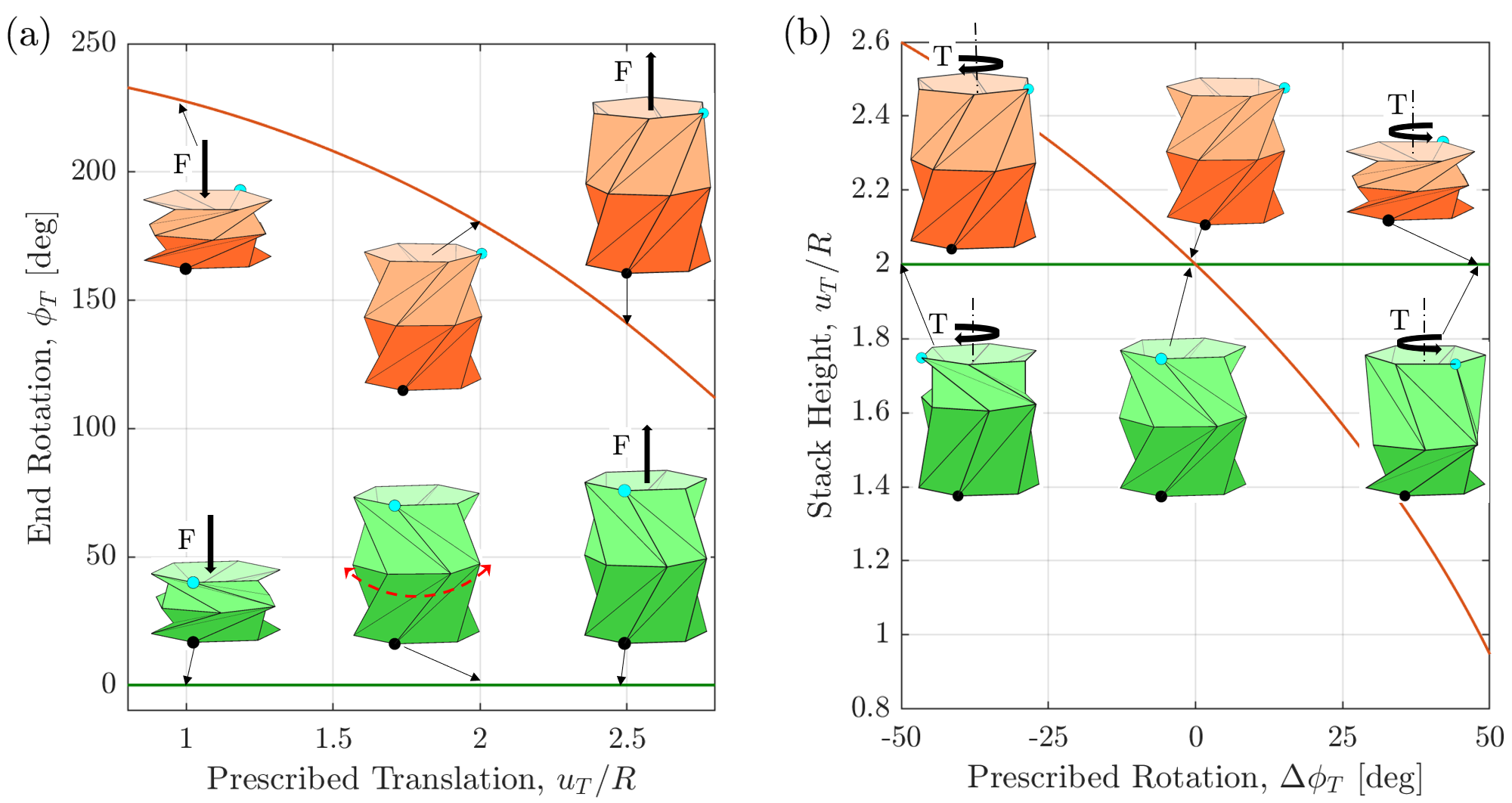}

                               \caption{Deformation of the top end of a pair of KOSPs. (a) Under uni-axial loading, and (b) under torsional loading. \textit{(Supplmentary video, S1 demonstrates this operation.)}} 

                               \label{KOSP_Eg}

               \end{figure*}

The decoupling of the motion at the free end can be observed by inspecting the $d$-KOSP in Fig. \ref{KOSP_Eg} (a) (Green).  When the bottom polygon is fixed while the upper end is subjected to a prescribed translational motion, $u_T$, the top polygon does not rotate as the height of the stack is increased (Note the circular blue marker placed on the top polygon does not rotate under the axial loading). The underlying kinematics of the KOSP allows the translational motion of the free end to occur free of rotation  by forcing the center polygon connecting both KOSs to undergo rotation and translation as the translation of the free end is taking place.  More specifically, the kinematics is such that the module whose crease orientation matches with the direction of the external rotation undergoes compression, while the other module undergoes expansion. On the other hand, when a prescribed translational motion, $u_T$, applied on the the top polygon of the $c$-KOSP (Orange), the top end undergoes coupled-rotational translational motion with extend range of operation as compared to a single KOS.

Similarly, as shown in Fig. \ref{KOSP_Eg} (b), when a prescribed rotational motion $\phi_T$, is applied at the top end of the $d$-KOSP, the connecting polygon  undergoes coupled translational-rotational motion that maintains that total height of the KOSP constant.  Videos demonstrating the different scenarios can be seen in supplementary video S1.

Another issue with the torsional behavior of a single KOS is that it is uni-directional. This is because the KOS is much stiffer when the applied torque opposes the folding direction of the panels than when it is applied in the same direction. As such, a single KOS always has an asymmetric restoring torque around its equilibrium state, which is not a desirable attribute. On the other hand, the $d$-KOSP is bi-directional and can be designed to have a symmetric restoring torque around its equilibrium, which is a key advantage over the single KOS.

It is therefore the goal of this paper to design and additively manufacture bi-directional tunable springs that have decoupled translational rotational degrees of freedom at their free end. The restoring force and torque behavior of those springs will be analyzed both numerically and experimentally using functional 3D-printed springs. The number of  equilibria, their stability, and bifurcations will also be analyzed as the precompression height of the $d$-KOSP is varied.

The rest of the paper is organized as follows: Section \ref{designfundamentals} introduces a simplified truss model, which can be used to study the qualitative quasi-static behavior of the KOS and uses it to analyze the equilibria of a single KOS. Section \ref{KOS modules} uses a truss model to investigate the quasi-static response behavior of $d$-KOSPs and analyzes the KOSPs possible equilibria and their bifurcations as the stack is pre-compressed to different heights. Section \ref{Exp Realization} presents an experimental study of the quasi-static behavior of the proposed $d$-KOSPs and illustrates that the responses are in qualitative agreement with the numerical findings.  Finally, section \ref{conclusion} presents the key conclusions.

\section{Restoring behavior of a single KOS}
\label{designfundamentals}

\begin{figure*}[htb!]
		\centering 
		\includegraphics[width=16cm]{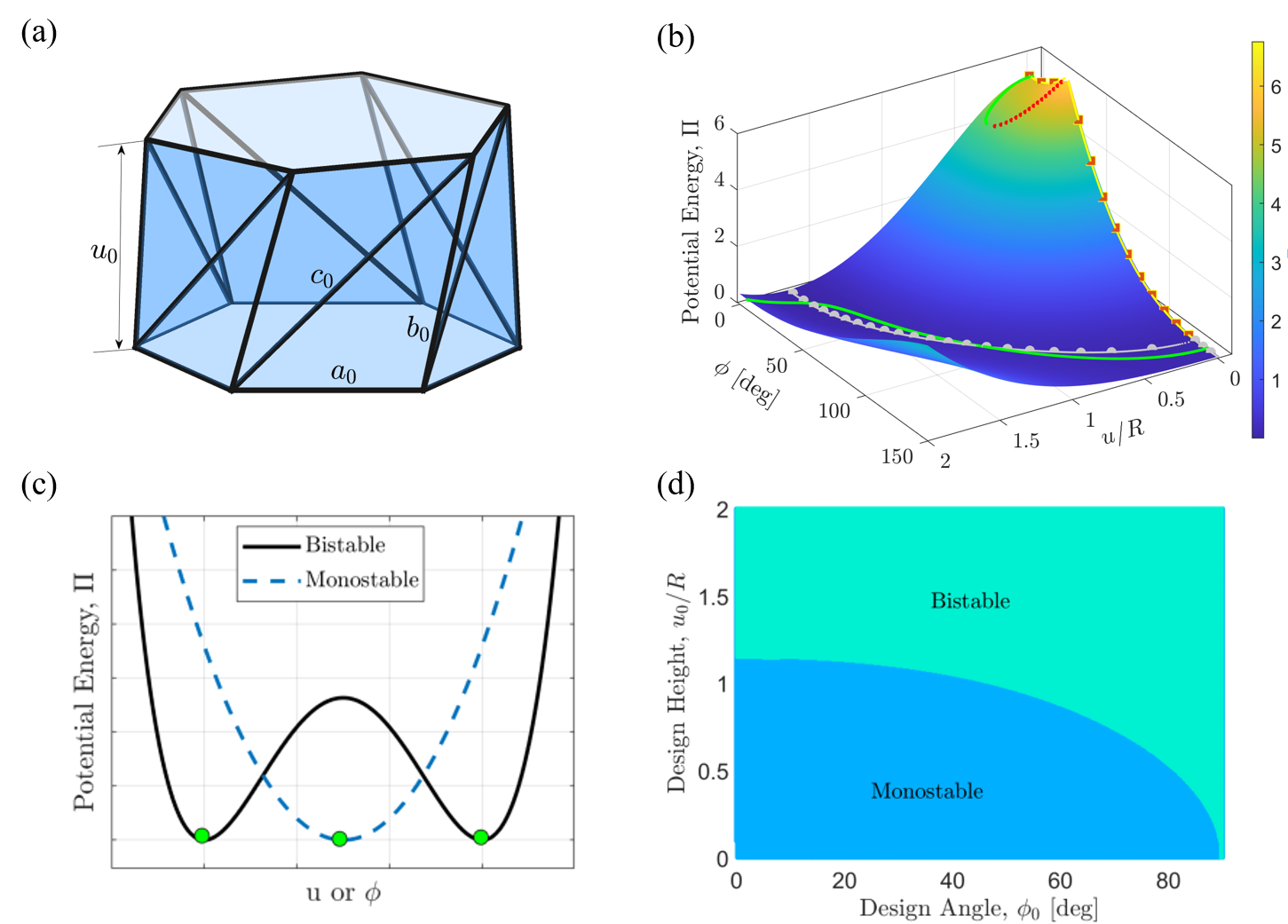}
		\caption{(a) Schematic representation of a truss model.  (b) Normalized potential energy function $\Pi$ using truss model of a KOS with design parameters, $u_0/R=1.65$, $\phi=15^o$ and $n=6$. (c) Typical potential energy plots- monostable and bistable. (d) Design map for $n=6$. } 
		\label{Scheme2}
	\end{figure*}

The restoring behavior (force and torque) of a KOS can vary greatly depending on the values of its geometric design parameters. In some cases, this can result in a single equilibrium configuration, while in others, there may be two equilibria. A qualitative understanding of the general behavior of KOSs can be obtained by using an axial truss model, in which each triangle in the KOS is represented by axially-deformable truss elements located at its edges\cite{YasudaKresling2017,RmasanaKIOS19}, as shown in Fig.~\ref{Scheme2} (a).\footnote{It is important to note that the truss model is only used in this paper as a qualitative guide for the choice of design parameters that result in different behaviors. More accurate, yet computationally expensive models can be found in \cite{DALAQ2022110541}.}

In the truss model, the relative position and orientation of the two end planes during deployment can be described by the length of the three edges of the triangle, $a$, $b$, and $c$,  in terms of the other design parameters as,
\begin{equation}
\begin{split}
a=2R\sin{\frac{\pi}n},\qquad
b=\sqrt{4R^2\sin^2\left({\frac{\phi-\frac{\pi}n}2}\right)+u^2},\qquad
c=\sqrt{4R^2\sin^2\left({\frac{\phi+\frac{\pi}n}2}\right)+u^2}. 
\label{edge_lengths}
\end{split}
\end{equation}
where $\phi$ and $u$ are, respectively, the relative angle and vertical distance between the end planes under loading. Assuming that the base of each triangle remains undeformed during deployment, and that the panels do not buckle under compression or encounter self-avoidance at small values of $u$, the total strain energy stored due to panel deformation can be approximated by
\begin{equation}
    \Pi=\frac{nEA}2\left[\frac{(b-b_0)^2}{b_0}+\frac{(c-c_0)^2}{c_0}\right],
    \label{PEnergy}
\end{equation}
where $EA$ is the axial rigidity of the truss elements. Here, $E$ is the elastic modulus, and $A$ is the cross-sectional area of the truss elements. Moving forward, all the results using the truss model are normalized with respect to the axial rigidity, or simply $EA$ is set to unity.

The equilibrium states ($u_e, \phi_e$) of the KOS are determined by minimizing the strain energy with respect to $u$ and $\phi$. Specifically, $\Pi_u|{(u_e,\phi_e)}= \Pi_\phi|{(u_e,\phi_e)}=0$ at any equilibrium state, where, $\Pi_u$ and $\Pi_\phi$ represents $\partial \Pi/\partial u$ and $\partial \Pi/\partial \phi$, respectively. An equilibrium configuration is considered physically stable only if it corresponds to a minimum in the strain energy, which is satisfied when $\Pi_{uu}\Pi_{\phi\phi}|{(u_e,\phi_e)}-\Pi_{u\phi}^2|{(u_e,\phi_e)}>0$ and $\Pi_{uu}>0$. 

Figure~\ref{Scheme2} (b) illustrates the normalized potential energy function, $\Pi$, for a KOS with the design parameters, $\phi_0=15^\circ$, $u_0/R=1.65$ and $n=6$. In the figure, the solid lines represent the condition $\Pi_\phi=0$ and $\Pi_{\phi\phi}>0$; i.e. the local minima lines, while the dotted lines represent the condition $\Pi_\phi=0$ and $\Pi_{\phi\phi}<0$; i.e., the local maxima lines. Similarly, the solid line with circular markers represents the condition $\Pi_u=0$ and $\Pi_{uu}>0$, and the line with square markers represent the condition $\Pi_u=0$ and $\Pi_{uu}<0$. These curves are important because they determine the route taken by the KOS during deployment. When the KOS is subjected to uni-axial loading without any external torque, the KOS follows the curve $\Pi_{\phi}=0$, whereas it follows the $\Pi_u=0$ curve when the KOS is subjected to a torque under no axial loading. Figure~\ref{Scheme2} (c) shows the typical potential energy functions of the KOS plotted across an independent variable, $u$ or $\phi$. The circular markers at the bottom of the potential energy curves represent the stable equilibria. 

The quasi-static behavior of the KOSs is highly dependent on the design parameters. Even a small variation in these parameters can lead to significant changes in the behavior of the KOS. The KOS is capable of exhibiting various qualitative restoring force characteristics. These include linear, nonlinear, and quasi-zero stiffness, among others. Depending on the number of stable equilibria that exist, KOSs are typically classified as mono-stable (one stable equilibrium) or bi-stable (two stable equilibria), see Fig.~\ref{Scheme2} (c). Figure \ref{Scheme2} (d) shows the design map that demarcates the design space ($u_0/R, \phi_0$) of the KOS into mono-stable and bi-stable regions for a KOS with $n=6$.

\section{Kresling Origami Pairs }
\label{KOS modules}
With the goal of expanding their utilizable space of application, we focus on understanding the quasi-static behavior of a pair of KOSs connected in series as shown previously in Fig.~\ref{KOSP_Eg}. The springs can be connected in two ways: either with the slope of the creases having the same sign  ($c$-KOSP) or with the slope of the creases having opposite signs ($d$-KOSP). When an external torque is applied to the top end of a $c$-KOSP while the other is fixed, the top end twists and compresses resulting in coupled translational-rotational motion. On the other hand, when the same load is applied to a $d$-KOSP, the top end only undergoes rotational motion without any translation effectively decoupling the translational from the rotational motions. 

To better understand the quasi-static behavior of the KOSP, we consider the truss model of a general, $N$-module KOS stack. It should be noted that the equations governing the mechanics of the truss model of the KOS with its assumptions are still applicable for the constituent KOSs. Having said that, the total strain energy stored in $N$-module stack of KOSs can be written as,

\begin{equation}
    \Pi_T=\sum_i^N \Pi_i = \sum_i^N\frac{n_iE_iA_i}2\left[\frac{(b_i-b_{i0})^2}{b_{i0}}+\frac{(c_i-c_{i0})^2}{c_{i0}}\right],
    \label{PEnergyT}
\end{equation}
where, 
\begin{equation}
\begin{split}
b_i=\sqrt{4R_i^2\sin^2\left({\frac{\phi_i-\frac{\pi}{n_i}}2}\right)+u_i^2},\qquad
c_i=\sqrt{4R_i^2\sin^2\left({\frac{\phi_i+\frac{\pi}{n_i}}2}\right)+u_i^2}.
\label{edge_lengthsT}
\end{split}
\end{equation}
Here, the subscript '$i$' refers to the different constituent KOSs in the stack.  The variables $\phi_i$ and $u_i$ are, respectively, the relative angle and the vertical distance between the end planes of the $i^{th}$ KOS module; $u_T=\Sigma u_{i}$ and $\phi_T=\Sigma \phi_i$, are, respectively,  the total height of the stack and net relative rotation between the stack's two ends; and finally, $\Pi_T$ is the potential energy of the stack. In this model, clockwise rotations are considered to be positive and vice versa. 

In response to any external loading along or about the longitudinal axis of the stack, the $N$ constituent KOS modules of the stack deform adjusting the $\phi_i$'s and $u_i$'s to counter balance the external loading with the net restoring force/torque. The  new arrangement of the KOS modules in response to the imposed external loads is such that it minimizes the total potential energy. The mathematical formulation of the optimization problem can be posed in the following way:
\begin{equation}
\begin{aligned}
& \underset{u_i,\phi_i}{\text{minimize}} \quad
  \Pi_T=\sum_{i=1}^{N} \Pi_i(u_i,\phi_i).\\
& \text{subject to} \quad \sum_{i=1}^{N} u_i = u_T, \qquad \sum_{i=1}^{N} \phi_i = \phi_T, \qquad u_{i}^{min}\leq u_i \leq u_{i}^{max}, \qquad \phi_{i}^{min}\leq \phi_i \leq \phi_{i}^{max} .  \\
\end{aligned}
\label{optim_truss}
\end{equation}
where $u_{i}^{min}$, $u_{i}^{max}$,$\phi_{i}^{min}$, $\phi_{i}^{max}$ are the minimum and maximum possible coordinates of the $i^{th}$ constituent. 

We use Matlab optimization tools to solve this problem, in which, at each iteration for a new set of ($u_T,\phi_T$), the program predicts the set of $\phi_i$'s and $u_i$'s that satisfies the constraints and evaluates the total potential energy, $\Pi_T$, using Equation~\ref{PEnergyT}.

	 	\begin{figure*}[htb!]
		\centering 
		\includegraphics[width=16cm]{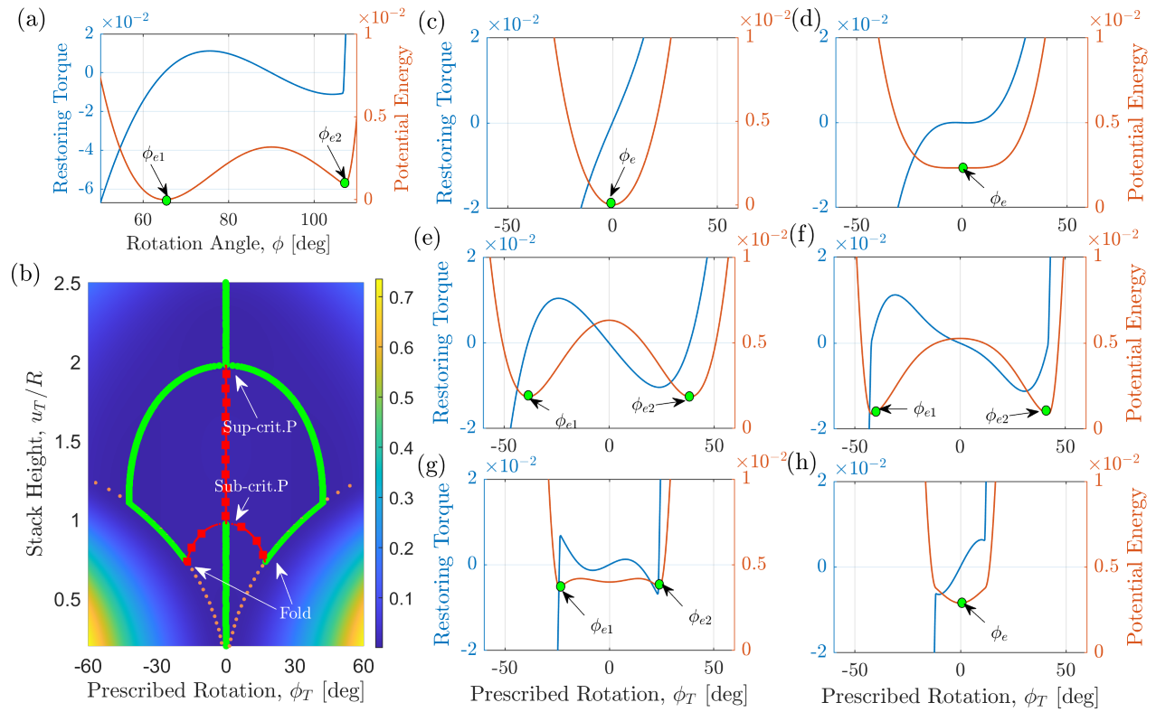}
		\caption{(a) Normalized restoring torque and normalized potential energy plots of a KOS module, $u_0/R=1.1$, $\phi_0=65^o$, and $n=6$. (b) The bifurcation diagram of the twin $d$-KOSP. Here, the solid green lines represent the stable equilibria, the red square markers represent the unstable equilibria and the dotted lines represent the folding limit of each KOS. (c), (d), (e), (f), (g) and (h) Normalized restoring torque and normalized potential energy plots of the $d$-KOSP for precompressed heights of (c) $u_{T}=2.2/R$, (d) $u_{T}/R=1.97$, (e) $u_{T}/R=1.5$, (f) $u_{T}/R=1.1$,  (g) $u_{T}/R=0.85$,  and (h) $u_{T}/R=0.6$.} 
		\label{Truss_BIF}
	\end{figure*}

\subsection{Restoring behavior of  $d$-KOSPs}
\label{OppKOSPs}
KOSPs formed by stacking KOSs with opposite orientation of their individual creases; i.e. $d$-KOSPs are of importance since, as aforedescribed, they offer bi-directional functionalities and decouple the motion at the free end. Thus, we dedicate this section to study their quasi-static restoring torque behavior, equilibria, and the bifurcation of those equilibria as the pre-compressed height of the stack, $u_T$, is varied. At each step of the analysis, the total height of the stack is changed and the potential energy function, restoring torque and equilibria of the KOSP are calculated using the algorithm described in Section \ref{KOS modules}. Figure~\ref{Truss_BIF} depicts such results for a twin $d$-KOSP consisting of two similar KOSs, each having the design parameters, $u_0/R=1.1$, $\phi_0=65^o$, and $n=6$, and the restoring torque behavior shown in  Fig.~\ref{Truss_BIF} (a). As can be clearly seen, the restoring torque of the single KOS is asymmetric bi-stable with uni-directional tendency.

As evident in the bifurcation diagram shown in Fig.~\ref{Truss_BIF} (b), the $d$-KOSP has a single equilibrium point at $\phi_T=0$ for pre-compressed heights $1.98<u_T/R<2.5$ (solid green line). Thus, the potential energy function is mono-stable and symmetric as shown in Fig.~\ref{Truss_BIF} (c) for $u_T/R=2.2$. The restoring torque of the $d$-KOSP is nearly linear, which is ideal for applications where linearity and symmetry under loading are key for performance. At $u_T/R=2$, the potential energy  becomes almost flat for any prescribed rotation near the equilibrium point, Fig.~\ref{Truss_BIF} (c). Thus, the stiffness becomes nearly zero around the equilibrium point resulting in a quasi-zero-stiffness (QZS) behavior. Such spring characteristics are ideal for  the design of broadband vibration absorbers and energy harvesters \cite{QZSBook,LIU2021121146}. Near $u_T/R=1.95$, the only stable equilibrium point of the $d$-KOSP loses stability through a super-critical pitchfork bifurcation (Sup-crit. P) and gives way to two stable equilibria on either side of the original equilibrium. Thus, the KOSP becomes of the symmetric bi-stable type. This is evident in the shape of the potential energy function shown in Fig.~\ref{Truss_BIF} (e) for $u_T/R=1.5$. It can be clearly seen that the potential energy function has two minima separated by a local maximum at $\phi_e=0$, which are characteristics of a symmetric bi-stable potential. The associated restoring force exhibits a negative stiffness at $\phi_e=0$ and positive stiffness for large values of $\phi_T$. Such bi-stable springs are key to the design of bi-stable mechanical switches and energy harvesters. 

As $u_{T}$ is decreased further, the two stable equilibrium branches diverge and the potential wells get deeper causing the magnitude of the negative local stiffness to increase. As such, it becomes more difficult to force the spring to move from one of its equilibria to the other.
At precisely, $u_{T}/R=1.1$, one of the KOSs in the stack becomes fully compressed, while the other is at its undeformed state. This is usually referred to in the literature as a self-contact point or as panel self-locking. The result is that the potential energy and the stiffness increase sharply and suddenly at these points as can be clearly seen in Fig.~\ref{Truss_BIF} (f). In essence, these points represent the limit of the $d$-KOSP operation. Any prescribed rotation of the $d$-KOSP beyond this point would only deform the panels; a process which requires high strain energy. Decreasing $u_{T}$ further below $u_{T}/R=1.1$, the non-trivial equilibria continue to follow the orange dotted curve which marks the self-contact points of the KOSP.

At about $u_{T}/R\approx 0.97$, the unstable equilibrium points represented by sqaure marker on Fig.~\ref{Truss_BIF} (b) regains stability through a sub-critical pitchfork bifurcation (Sub-crit. P), and the KOSP becomes tri-stable as can be seen in Fig.~\ref{Truss_BIF} (g). The KOSP remains tri-stable for a very short range of $u_{T}$, before the two non-zero stable solutions collide with the unstable solution and destruct each other in a fold bifurcation (Fold) at $u_{T}/R\approx 0.74$. Beyond this point, the KOSP becomes mono-stable again with the trivial position, $\phi_e=0$, being the only equilibrium point, as can be seen in Fig.~\ref{Truss_BIF} (h).

	 	\begin{figure*}[htb!]
		\centering 
		\includegraphics[width=16cm]{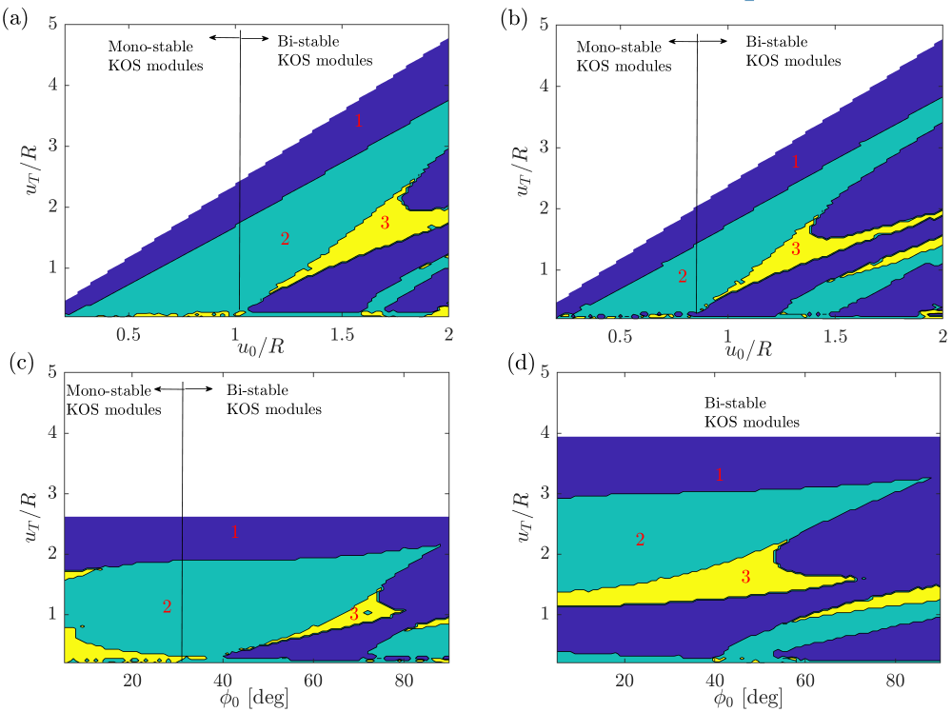}
		\caption{Influence of the KOS module design parameters on the number of stable equilibria of the twin $d$-KOSP, (a) $\phi_0=45^o$, (b) $\phi_0=65^o$, (c) $u_0/R=1.1 $ and (d) $u_0/R=1.65$. } 
		\label{Truss_regions}
	\end{figure*}

	 	\begin{figure*}[htb!]
		\centering 
		\includegraphics[width=16cm]{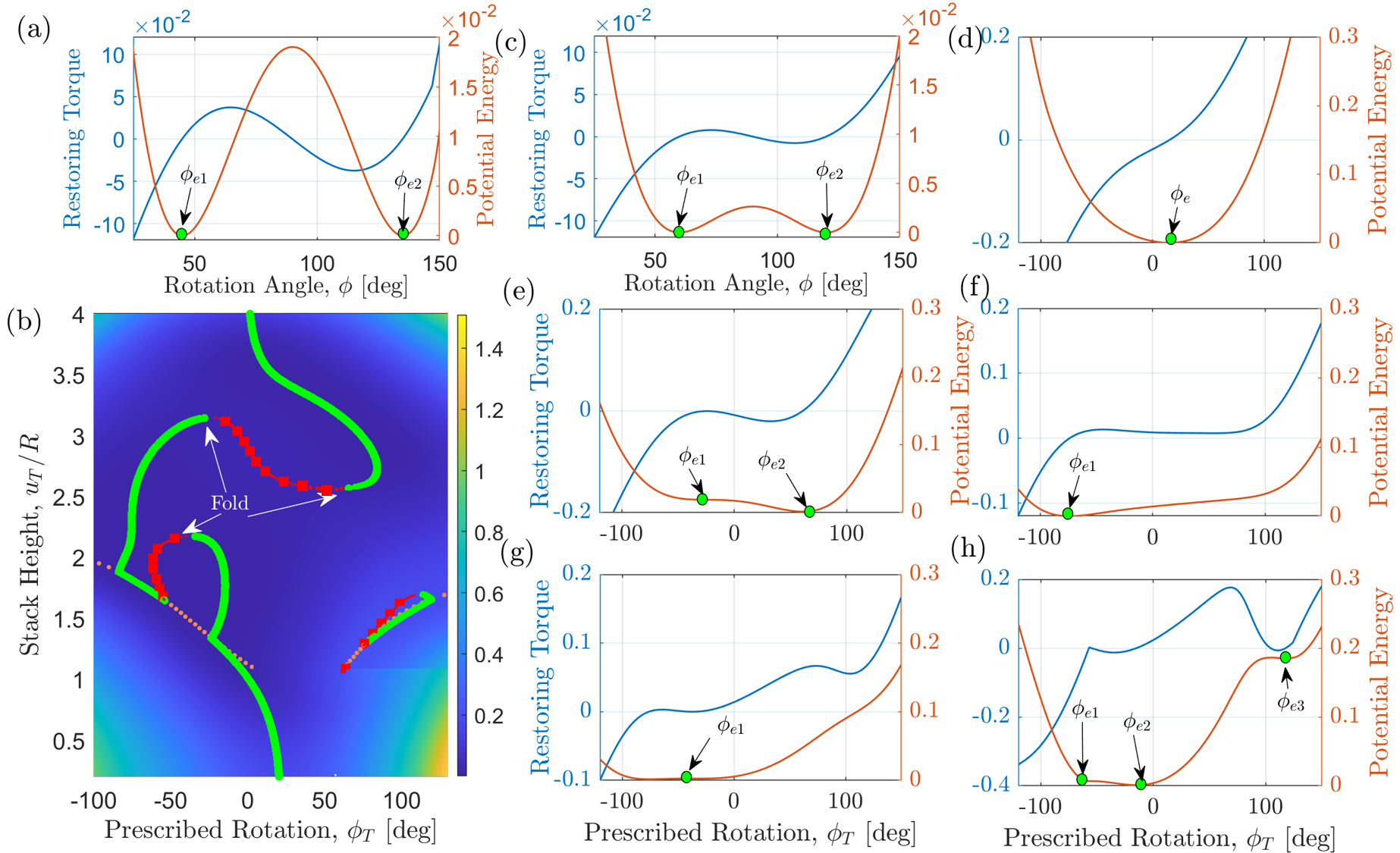}
		\caption{(a) Normalized restoring torque  and normalized potential energy plots of  KOS1: $u_0/R=1.65$, $\phi_0=45^o$, and $n=6$. (b) Bifurcation diagram of the $d$-KOSP (solid green line represents stable solutions while square markers represent unstable solutions). (c) Normalized restoring torque and normalized potential energy plots of  KOS2: $u_0/R=1.875$, $\phi_0=60^o$, and $n=6$. (d), (e), (f), (g) and (h) Normalized restoring torque and normalized potential energy plots of the $d$-KOSP for precompressed heights of (d) $u_{T}/R=3.525$, (e) $u_{T}/R=3.152$, (f) $u_{T}/R=2.50$,  (g) $u_{T}/R=2.17$,  and (h) $u_{T}/R=1.67$. } 
		\label{Truss_BIF_NT}
	\end{figure*}

In Fig.~\ref{Truss_regions} (a,b), we generalize the bifurcation diagram shown in Fig.~\ref{Truss_BIF} (b) into bifurcation maps that demarcate the design space of $u_T/R$ and $u_0/R$ into different domains based on the number of stable equilibria that the twin $d$-KOSP possesses; i.e., which design parameters lead to mono-stable behavior, and which ones lead to a bi- or tri-stable behavior. The different colored regions with numbered labels represent the different number (as labeled in the figure) of the stable equilibria in that configuration. Those maps are generated for modules with two different values of $\phi_0$, namely $\phi_0=45^o$ and $\phi_0=65^o$. For the most part, we can see that the mono-stable behavior is the easiest to realize followed by the bi-stable behavior, then the tri-stable one, and that the region of design parameters leading to the tri-stable behavior shrinks when $\phi$ is increased.  Moreover, in both of the cases, tri-stability is very difficult to achieve when the KOS modules are mono-stable as compared to when the KOSP is designed using bi-stable KOS modules. Figure~\ref{Truss_regions} (c) and (d), show similar maps for $u_0/R= 1.1 $ and $1.65$, respectively, with $\phi_0$ being the bifurcation parameters. It can be clearly seen that larger values of $u_0/R$ allow for larger regions in the design space to construct bi- and tri-stable KOSPs. One interesting observation resulting from the numerical analysis is that the $d$-KOSP can be designed to become mono-stable, bi-stable or even tri-stable irrespective of the type of the stability of its constituents.

The symmetric response of the twin module $d$-KOSP is a feature that is most often desired in designing springs, but is not a constraint, if otherwise, an asymmetric restoring force response is desired. Asymmetry can be easily achieved by constructing the $d$-KOSP using two different KOSs. For instance, in Fig.~\ref{Truss_BIF_NT} (b), we plot the bifurcation diagram for a $d$-KOSP constructed by combining the two different bi-stable KOSs whose restoring behavior is shown in Figs.~\ref{Truss_BIF_NT} (a,c); namely,  KOS1: $u_0/R=1.65$, $\phi_0=45^o$, and $n=6$ and, KOS2: $u_0/R=1.875$, $\phi_0=60^o$, and $n=6$. 

A first glance reveals that the bifurcation diagram is more complex and is no longer symmetric around $\phi_e=0$.  At the uncompressed height; i.e. $u_{T}/R=1.65+1.875=3.525$, there is a net offset in the equilibrium rotation angle which is $60^o-45^o=15^o$, relative to the other end of the KOSP.  The potential energy function is mono-stable despite both constituents being bi-stable and the restoring force is of the nonlinear hardening type, as shown in Fig.~\ref{Truss_BIF_NT} (d).

When the KOSP is pre-compressed, the force deforms the two springs differently since they have different stiffnesses. The softer spring, here KOS2,  undergoes compression and rotation first under the applied load. In the process of compression, KOS2, gains stiffness up to the point $u_{T}/R=3.2$, where it becomes stiffer than KOS1. At this point, KOS1 starts to deform and a new equilibrium point is born causing the potential energy to become bi-stable and asymmetric with two equilibrium angles occurring at $\phi_{e1}=-28^o$, and, $\phi_{e2}=62^o$; see Fig. ~\ref{Truss_BIF_NT} (e) obtained at $u_T/R=3.152$. 

The bi-stable asymmetric behavior continues to persist down to $u_T/R \approx 2.5 $, where the spring behavior becomes nearly of the asymmetric quasi-zero stiffness  type; see Fig. ~\ref{Truss_BIF_NT} (f). Subsequently, the behavior of the springs becomes very complex as shown in Fig. ~\ref{Truss_BIF_NT} (g) and (h) for $u_T/R=2.17$ and $u_T/R=1.67$, respectively. The bifurcation diagram also reveals that the tri-stable behavior cannot be achieved using this combination of spring modules.

\section{Experiments}
\label{Exp Realization}
\subsection{Fabrication}
\begin{figure*}[htb!]
		\centering 
		\includegraphics[width=15.0cm]{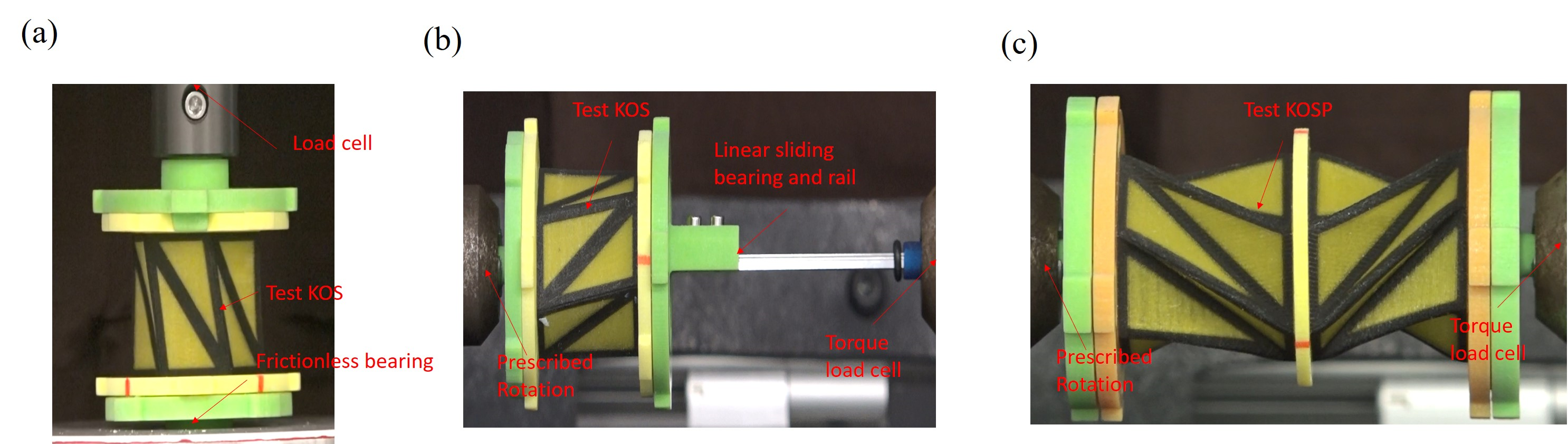}
		\caption{Experimental setup for uni-axial testing of KOS. (b) Experimental setup for torsional testing of KOS. (c) Experimental setup for torsional testing of KOSP.}
		\label{3DKOSfab}
	\end{figure*}

Numerical simulations have revealed that the modularity of the KOS can be used to construct functional KOSPs with unique and made-to-order restoring characteristics. The operating range can be increased and the restoring force/torque further tuned by stacking a larger number of unit springs $N\geq2$. To employ these desirable characteristics in a realistic environment, such springs must be durable, and their manufacturing process be systematic and repeatable. Thus, relying on paper folding is obviously not the optimal approach. In a recent article \cite{KHAZAALEH2022109811}, we used 3D printing to produce KOS modules, demonstrating exceptional functionality, repeatability, and high durability.  In the proposed design, the basic triangles of each KOS were modified to allow for easy folding and stretching at the panel junctions while still retaining enough stiffness to conform to the Kresling origami pattern and withstand loading. The fabrication process used the Stratasys J750 3D printer with the polyjet method, which utilized two different materials for each panel. The central rigid core of each panel was made of a rigid plastic polyjet material called \textit{Vero}, while the outer frame was made of a flexible rubber-like polyjet material called \textit{TangoBlackPlus}. The flexibility of the outer frame enables folding and stretching at the interfaces. Fig.~\ref{3DKOSfab} (a) shows an example of the fabricated KOS module. The new design further introduced additional geometric parameters, namely the width of the flexible material, $w$ and the thickness of the panels, $t$. These are important parameters that provide additional freedom in designing the KOS modules.  Each KOS is reinforced using two end plates with circular holes that are concentric with the longitudinal axis of the KOS. These plates are added to increase the stability of the KOS, and to prevent damages under unwarranted non-axial loads. The holes allow air to escape during deployment.  Using the manufacturing approach proposed in our previous article \cite{KHAZAALEH2022109811},  we construct KOSPs for experimental testing. Similar and/or different  KOSs are used interchangeably to form various stack combinations, and different crease orientations.

\subsection{Experimental Testing}
\label{Experiment}
The experimental portion of this study involves two phases: axial testing and torsional testing. The axial tests are performed to determine the restoring force behavior of the KOS under compressive and tensile loads. These tests are conducted using an Instron Dual Column 5960 universal testing machine. A controlled fixed rate displacement of 0.2 mm/s is applied to the top end of the KOS while the bottom end is placed on a specially designed platform that can freely rotate about a common centroidal axis, as shown in Fig.~\ref{3DKOSfab} (a). The restoring force is measured using a load cell, and the rotation of the bottom end is tracked using digital image correlation (DIC) tools.

During the torsional tests, the KOS is subjected to a controlled rate of cyclical torque, including clockwise and anticlockwise rotations, using an Instron MicroTorsion MT1 machine, as depicted in Fig.~\ref{3DKOSfab} (b). One end of the KOS is clamped using a chuck that is connected to a motor, while the other end is placed on a sliding bearing that allows longitudinal motion to be free while the rotary motion is restrained. For the torsional tests of the KOSPs, we require the total length of the KOSP, $u_T$, to be fixed. Thus, we remove the sliding bearing and prevent it from rotating or translating. A torque cell is used to measure the applied torque at the fixed end, and the relative longitudinal displacement is monitored using DIC tools. It is important to note that the results presented in this work are based on a controlled rotational rate of $20^{o}/min$, although other rotational rates ranging from $10^{o}/min$ to $100^{o}/min$ were also tested, but the effect of the rate on the response was found to be negligible. Furthermore, the entire test setup is placed in the horizontal plane to eliminate the influence of gravity on the quasi-static responses, which is crucial since the KOS is free to slide during  axial testing.

It is important to mention that five samples with the same design parameters are tested in both the axial and torsional testing of the KOS modules, and the average of the responses under compression, tension, clockwise and anticlockwise rotation are recorded. The total potential energy is calculated by integrating the measured restoring force across the prescribed displacement in the case of uni-axial testing and integrating the measured torque over the prescribed rotation in the case of torsional testing. 

\subsection{Experimental Results}
\label{ExpResults}
\begin{figure*}[htb!]
		\centering 
		\includegraphics[width=14cm]{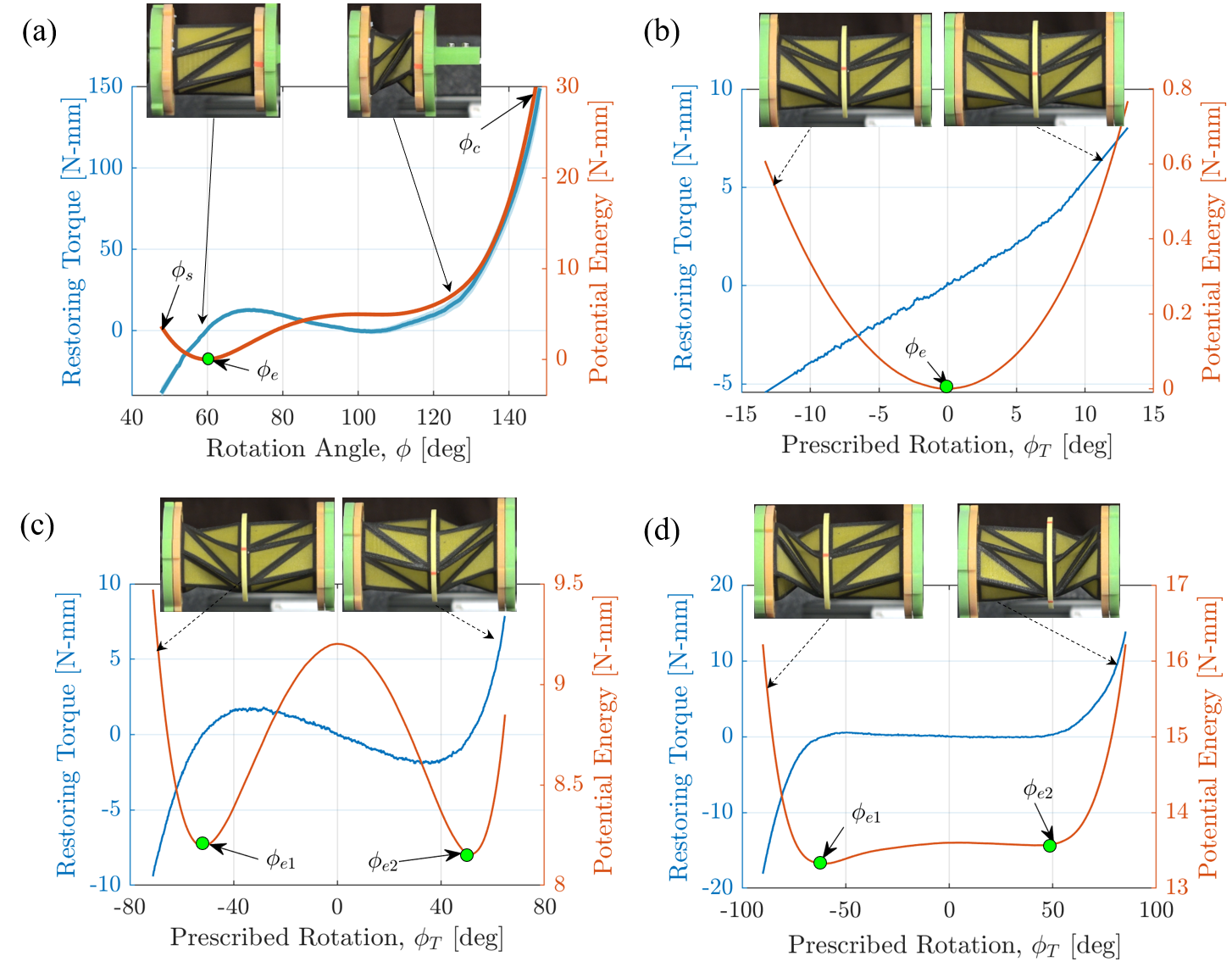}
		\caption{(a) Experimental restoring torque and the associated potential energy plots under torsion tests of a single KOS module with design parameters: $u_0/R=1.875$, $\phi_0=60^o$, $n=6$. (b,c,d ) Restoring torque and potential energy plots of a twin $d$-KOSP with (d) $u_{T}/R=3.75$, (e) $u_{T}/R=3.36$, and (f) $u_{T}/R=3.05$.} 
		\label{ExpKOSP}
	\end{figure*}

We start by testing the quasi-static torsional behavior of a KOS with the geometric parameters $u_0/R=1.875$, $\phi_0=60^o$, $R=15$mm,  $n=6$, $t=0.75$ mm and $w=1.5$ mm as depicted in Fig.~\ref{ExpKOSP} (a).  Our goal is to first understand the behavior of the unit cell forming the KOSP. To achieve this objective, we tested the KOS samples using the Instron torsion testing machine, which was configured to maintain zero axial loading on the KOSs throughout the test. During the test, we prescribed the rotation angle, $\phi$, at one end of the module and recorded both the torque and the instantaneous height of the structure, $u$. Positive rotation caused compression in the KOS module, while negative rotation resulted in expansion. To prevent the module from suffering permanent deformation or damage, we determined the limits of rotation, including the stretch limit and compression limit. Here, the stretch limit, $\phi_s$, refers to the point at which $\partial u/\partial \phi$ becomes large in uni-axial testing, while the compression limit, $\phi_c$, is the point before delamination begins to occur. For the considered KOS, $\phi_s=48.5^o$ while $\phi_c=147.5^o$ clearly demonstrating one of the key disadvantages of the single KOS design, which lies in the fact that it reaches its stretch limit much faster than its compression limit due to its inherent kinematics.

 Figure~\ref{ExpKOSP} (a) illustrates the restoring torque and the calculated potential energy function for this KOS. It is evident that the restoring force is asymmetric with the KOS exhibiting a single equilibrium point at the undeformed state $\phi_{e}=60^o$, where the restoring torque is zero. As the angle is increased,  the restoring torque monotonously increases up to  $\phi=71^o$, after which it starts to decrease resulting in negative torsional stiffness up to $\phi=89^o$. Thereafter, the KOS loses much of its load carrying ability and the potential energy forms a plateau which extends up to $\phi=110^o$. Note that the restoring torque approaches but never crosses zero, thus there is no other equilibrium point and the spring is mono-stable. On rotating further, the KOS begins to get stiffer and the triangular panels begin to interact and avoid each other near $\phi=130^o$. On the other hand, upon expansion the KOS from its undeformed state, the stiffness rapidly increases and the KOS quickly reaches its stretch limit of  $\phi_s=48.5^o$.
 
Next, we investigate the restoring behavior of a twin $d$-KOSP constructed using a pair of the tested KOS, $u_T/R=2\times1.875=3.75$. Following the testing procedure described in Section~\ref{Experiment}, the restoring torque is measured as shown in Fig.~\ref{ExpKOSP} (b). It is clearly evident that, for this value of $u_T/R$, the restoring torque of the spring is nearly symmetric and linear around $\phi=0$ despite each KOS forming the stack being highly asymmetric and nonlinear. When the spring is precompressed to a height of $u_T/R=3.36$, the spring becomes bi-stable with two equilibrium points ($\phi_{e1}=-53^o$  and  $\phi_{e2}=52^o$) as shown in Fig.~\ref{ExpKOSP} (c). The restoring force is bi-stable and nearly symmetric despite the constituents being highly asymmetric and mono-stable. Further decrease of the height to $u_T/R=3.05$ decreases the depth of the potential wells, and the restoring force becomes nearly of the QZS type as shown in Fig.~\ref{ExpKOSP} (d).

\begin{figure*}[htb!]
		\centering 
		\includegraphics[width=16cm]{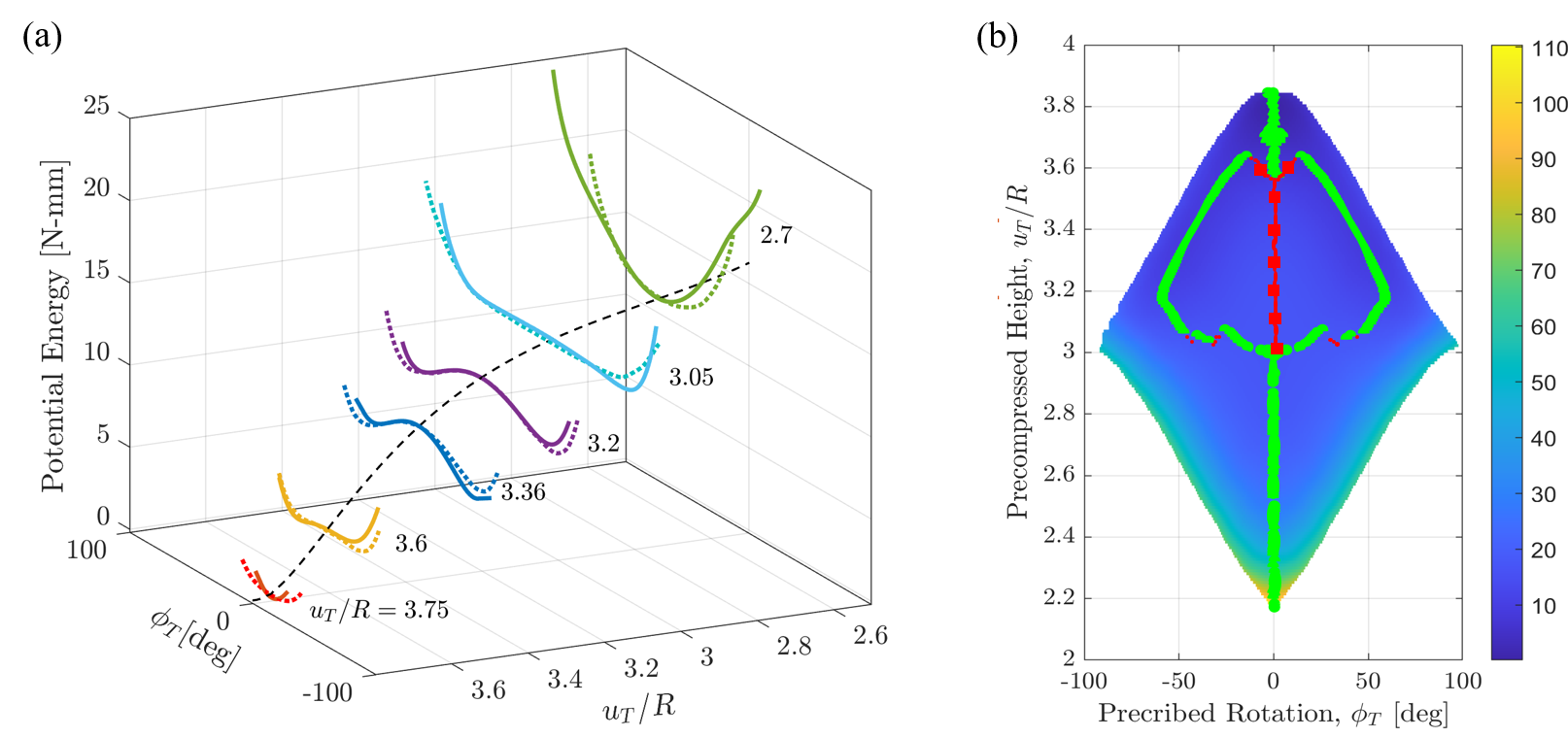}
		\caption{(a) Potential energy function of the KOSP at different precompressed heights. Solid lines represent experimental results, dotted lines represent simulated results. (b) Bifurcation diagram of  the twin module $d$-KOSP simulated using the experimental data from Fig.~\ref{ExpKOSP}(a). Solid green lines represent the stable equilibria, while the square markers represent the unstable ones. The contour map represents the potential energy of the stack measured in N-mm. } 
		\label{ExpKOSP2}
	\end{figure*}

To generate the full bifurcation diagram for different values of $u_T/R$ without having to repeat this experiment a large number of times, we use a numero-experimental interpolation approach. The approach employs an algorithm similar to the one described in the numerical analysis, Equation~\ref{optim_truss}, to minimize the change in the total potential energy of the KOSP structure during its operation. However, here instead of evaluating the potential energy at the iteration variables ($u_T,\phi_T$), we interpolate the potential energy of the modules from the experimental data used in Fig.~\ref{ExpKOSP2} (a). As with the simulation of truss KOSPs, we initially set the length of the stack, $u_T$ to a certain value and then iteratively change the rotation angle, $\phi_T$. Then the algorithm makes an initial calculated guess (feasible values) of $\phi_1$ and $\phi_2$, such that they are bounded within the stretch and compression limits of the modules and also satisfy $\phi_1+\phi_2=\phi_T$. Using the experimental data, the program evaluates $u_1$, and $u_2$ and verifies that $u_1+u_2=u_T$. Once it is verified, the program interpolates $\Pi_1$ and $\Pi_2$ for $\phi_1$ and $\phi_2$ and evaluates the net potential energy $\Pi_T$. The optimization tool then uses fmincon in Matlab to iteratively searche for the minimum of $\Pi_T$ by moving towards the feasible region of the objective function. 

The potential energy functions of the KOSP  obtained using the algorithm are compared to their experimental counterparts as shown in Fig.~\ref{ExpKOSP2} (a) for different values of $u_T/R$. Here, the solid lines represent the actual experimental findings while the dotted lines represent those attained using the aforedescribed algorithm. It can be clearly seen that there is a general qualitative agreement, which could be used to predict an experimental bifurcation diagram of the KOSP as a function of $u_T/R$. This diagram is shown in Fig.~\ref{ExpKOSP2} (b) demonstrating the ranges of $u_T/R$ for which the KOSP has a mono- versus bi- or even tri-stable behavior. The surface plot shows the potential energy (strain energy), and the stable and unstable equilibria are represented by a green line and red colored square markers, respectively. The white region represents the values of $u_T$ and $\phi_T$ that lie outside the operation range of the KOSP; i.e. the stretch and compression limits of the two constituent modules. Generally, we can see that  a $d$-KOSP constructed from an identical pair of KOSs possesses a symmetric potential function  that allows for a similar response to both clockwise and counter-clockwise rotations, which is not possible with a single KOS.  Additionally, $d$-KOSPs exhibit an increased range of operability in both rotation and longitudinal deployment. The stiffness of $d$-KOSPs can be tuned or programmed, including quasi-zero stiffness, providing great control over their behavior for specific applications.

	 	\begin{figure*}[htb!]
		\centering 
		\includegraphics[width=16.5cm]{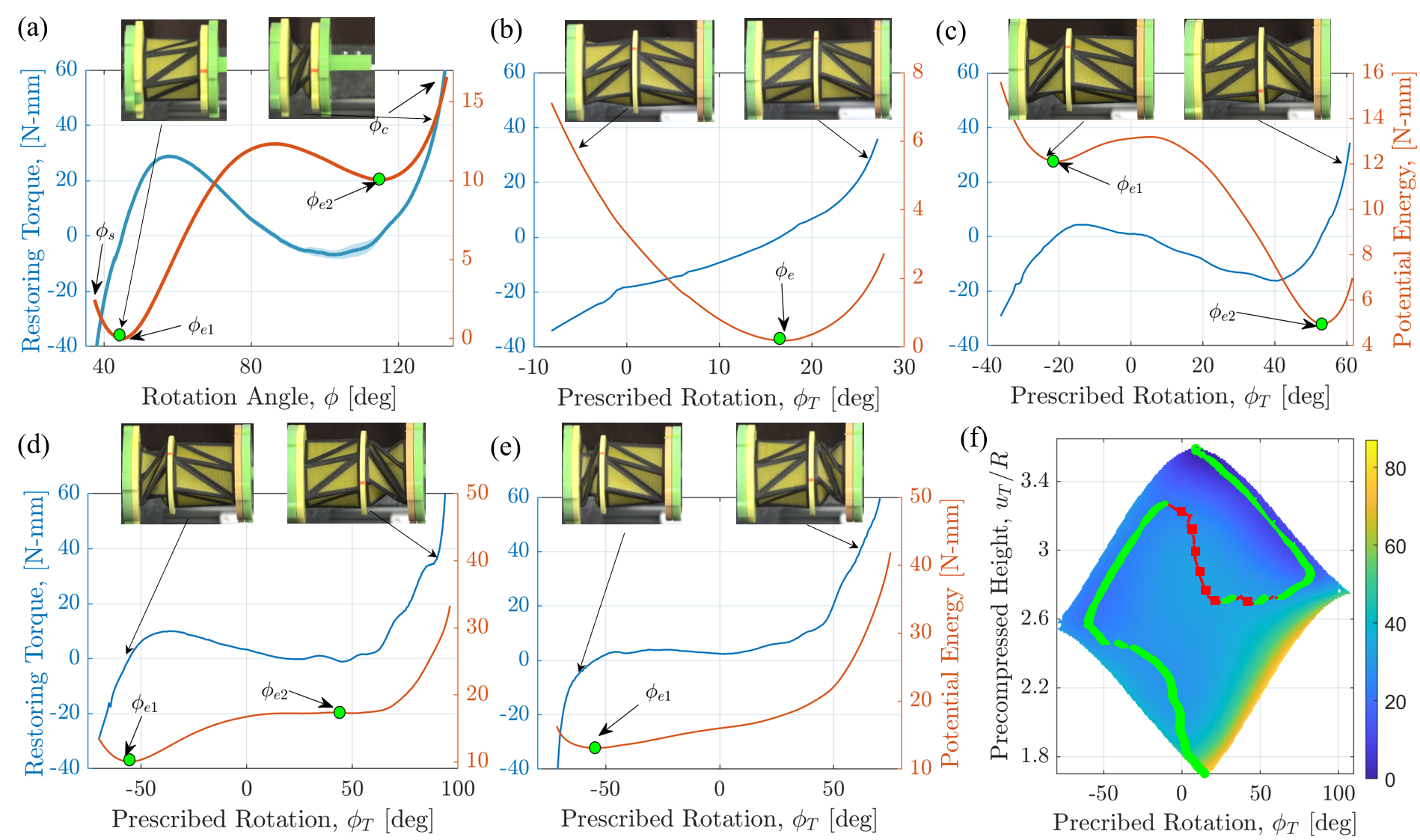}
		\caption{(a) Restoring torque and the potential energy plots under torsion tests (experimental) of a KOS module with design parameters, KOS2:: $u_0/R=1.65$, $\phi_0=45^o$, $n=6$; (b,c,d,e ) Restoring torque and potential energy plots (experimental) of the KOSP at precompressed heights (b) $u_{T}/R=3.525$, (c) $u_{T}/R=3.2$, (d) $u_{T}/R=2.7$, (e) $u_{T}/R=2.5$; (f) Bifurcation of the $d$-KOSP. Solid green lines represent the stable equilibria, while the square markers represent the unstable ones. The contour map represents the potential energy of the stack measured in N-mm.}
		\label{ExpKOSP_mix}
	\end{figure*}
	
Figure~\ref{ExpKOSP_mix} shows the quasi-static response behavior for a KOSP consisting of two different KOSs. KOS 1 is the mono-stable spring studied in Fig.~\ref{ExpKOSP}, while KOS 2 is bi-stable with  $u_0/R=1.65$, $\phi_0=45^o$, $R=15$ mm and $n=6$. The restoring torque and potential energy of KOS2 are as shown in Fig.~\ref{ExpKOSP_mix} (a). The two equilibria of KOS 2 occur at $\phi_{e1}=45^o$ and $ \phi_{e2}= 116^o$. The $d$-KOSP constructed using those springs is experimentally tested at different values of $u_T/R$ and the restoring torque responses are recorded. Fig.~\ref{ExpKOSP_mix} (b) shows those responses for the uncompressed height $u_T/R=3.525$ where the potential energy function is mono-stable but asymmetric with a single equilibrium point occurring near $\phi_e= 15^o$. The restoring force is weakly nonlinear with a slight hardening behavior near the stretch and compression limits. When the KOSP is precompressed to a height of $u_T/R=3.2$, the KOSP becomes bi-stable asymmetric with the left potential well being much deeper than the right one. The right potential well becomes shallower as the KOSP is precompressed further up to the point where the second minimum in the potential energy function disappears near $u_T/R= 2.7$, Fig.~\ref{ExpKOSP_mix} (d). In this case, the restoring force becomes nearly QZS around $\phi_T=25^o$. As the KOSP is precompressed further to $u_T/R= 2.5$, the spring becomes of the QZS type but with asymmetric characteristics Fig.~\ref{ExpKOSP_mix} (e).

Figure~\ref{ExpKOSP_mix} (f) depicts the interpolated bifurcation diagram which reveals that the $d$-KOSP will always exhibit an asymmetric response with the characteristics being of the mono-stable type for large and small precompression heights and bi-stable for intermediate ones. 

\section{Conclusion}
\label{conclusion}
This paper focuses on the use of serially connected Kresling Origami Springs (KOS) to design bi-directional programmable springs that offer decoupled translational and rotational degrees of freedom. The behavior of these springs is investigated both numerically, using a truss model, and experimentally, using functional 3D printed springs. The study reveals that, by varying the precompressed height of the KOSPs, interesting bifurcations of the static equilibria emerge leading to mono-, bi-, tri-stable, and QZS restoring elements with either symmetric or asymmetric restoring behavior. This is unlike single KOSs whose translational and rotational degrees of freedom are always coupled and cannot be designed to have a tri-stable behavior or to possess a symmetric restoring force behavior.  The availability of such springs open up new avenues for developing restoring elements with programmable responses to external stimuli, which could lead to innovative applications in various fields, such as robotics and energy storage \cite{YasudaKresling2017,MasanaKIMS}. The results of this study contribute to the expanding body of knowledge on the potential use of origami principles for engineering applications. The findings also provide new insights into the behavior of KOSPs and offer a foundation for future research in developing multifunctional materials using the Kresling origami pattern.
	
\section*{Acknowledgements}

We thank Core Technology Platforms at New York University Abu Dhabi (NYUAD) for providing their resources and support during fabrication and testing of the KOSs. Parts of this work was supported by the NYU-AD Center for Smart Engineering Materials which is under the full support of Tamkeen under NYUAD RRC Grant No. CG011.

\section*{Data Availability}
The data that support the findings of this study are available from the corresponding author upon reasonable request.

\bibliography{KOS}

\end{document}